\documentclass[12pt,english]{article}
\pdfoutput=1
\usepackage{setspace}
\usepackage{dcolumn}
\usepackage{bm}
\usepackage{amsmath, amssymb, amsthm, mathrsfs}
\usepackage{graphicx}
\usepackage{listings}
\usepackage[T1]{fontenc}
\usepackage{cite}

\usepackage[unicode=true,bookmarks=true,
bookmarksnumbered=false,bookmarksopen=false,
breaklinks=false,
pdfborder={0 0 1},
backref=false,colorlinks=true]{hyperref}

\newcommand{\bra}[1]{\langle #1 |}

\usepackage{xcolor}
\newcommand{\be}{\begin{equation}}
\newcommand{\ee}{\end{equation}}

\newcommand{\autocite}{\cite}

\begin{document}
\baselineskip=16pt

\vspace{1cm}
\thispagestyle{empty}
\begin{center}
{\LARGE\bf
What Emergence Can Possibly Mean
}\\
\bigskip\vspace{0.5cm}{
{\large Sean M. Carroll,\textsuperscript{1} and Achyth Parola\textsuperscript{2}\vspace{1mm}}
} \\
 {\it \textsuperscript{1}\mbox{Johns Hopkins University and Santa Fe Institute, seancarroll@gmail.com}\vspace{1mm}
\textsuperscript{2}\mbox{Johns Hopkins University, aparola1@jhu.edu}\vspace{1mm}
}
 \end{center}
\bigskip
\centerline{\large\bf Abstract}

\begin{quote} \small
    We consider emergence from the perspective of dynamics: states of a system evolving with time.
    We focus on the role of a decomposition of wholes into parts, and attempt to characterize relationships between levels without reference to whether higher-level properties are ``novel'' or ``unexpected." 
    We offer a classification of different varieties of emergence, with and without new ontological elements at higher levels.

    Submitted to a volume on Real Patterns (Tyler Milhouse, ed.), to be published by MIT Press.
\end{quote}

\section*{Introduction}

Everyone agrees that emergence is important, but they don't agree on what the word should mean. 
The basic idea is that, when a system can be thought of as a composite of many parts, there can be novel properties or behaviors of the composite system that are not readily predictable from considerations of the properties and behaviors of the parts themselves \cite{gibb2019routledge,wilson2021metaphysical}.
Emergence is relevant when there is both a ``micro'' (more fundamental, comprehensive, lower-level) description and a ``macro'' (emergent, coarse-grained, higher-level) description of the same system, both of which capture important properties.

Emergence is ubiquitous, and is crucial to how we deal with reality: we are able to model and predict features of the world with dramatically incomplete information about it.
We can talk fruitfully about tables and chairs, not to mention people, without anything close to a complete picture of the elementary particles and forces out of which they are apparently made.
Properties like ``temperature" or ``wetness'' or ``irritability" do not apply to individual particles, but are useful emergent features of the macroscopic world. Even more abstract notions like chance and probability may be thought of as emergent phenomena. \autocite{List2015-LISEC}

Ambiguities arise when we try to pin down the precise sense in which emergent properties are ``novel."
There are different conceptions of emergence, which might be relevant to different circumstances, but which also might require very different ontological frameworks.
It is traditional to distinguish between the ideas of ``weak'' emergence, where higher-level properties follow in principle from lower-level ones, and ``strong'' emergence, where higher-level properties are truly new \autocite{sep-properties-emergent}.
Strong emergence is an especially popular concept in some approaches to understanding consciousness, but it is also invoked in discussions of the origin of life and elsewhere \autocite{Chalmers2006-CHASAW, Tononi2016,Sharma2023}.
There is also a distinction between ``epistemic'' emergence, which refers to an ability to capture features of systems depending on different levels of knowledge about them, and ``ontological'' emergence, where emergent properties are thought to really exist in a way that is not reducible to lower-level properties.

While these distinctions can be useful, they remain imprecise, and that imprecision opens the door to unresolved conceptual issues. 
Perhaps the most notable is the relationship between strong or ontological conceptions of emergence and the scope and success of fundamental physics.
Granting that there is a lower-level theory that is successful in some domain of applicability, and that the macroscopic system is composed of lower-level parts, it seems strange to accept that success and nevertheless propose that a successful higher-level theory might not be deducible, even in principle, from the lower-level one.
Wouldn't that simply be to say that the lower-level theory is wrong, or at least incomplete, and should be appropriately modified to apply even in macroscopic situations?
A human brain, for example, contains electrons, quarks, photons, and other particles and fields described very accurately by the rules of quantum field theory (QFT) \autocite{carroll2022quantum}.
Given any particular state of those ingredients, the theory makes clear predictions for how that state will evolve over time.
To believe that consciousness is strongly emergent would require that the predictions of this QFT, for example for the behavior of electrons, are incorrect when applied to a situation like a human brain \autocite{carroll2021consciousness}.
That is certainly conceivable, but it seems hard to imagine how the general principles of QFT fail to apply in this situation, or what the actually correct theory would look like.
This difficulty makes it important for strong emergentists to be extremely precise about what they mean, especially in relation to fundamental physics.

Beyond that, there is something vague about the novelty component of emergence. 
Emergent properties are variously described as unexpected, surprising, or impossible to derive, given only the lower-level description \autocite{Bedau1997-BEDWE}.
But whether or not something is unexpected or novel seems to be a matter of individual judgment, rather than a rigorous designation. 
It is even hard to be perfectly precise about when something is impossible to derive; maybe it simply hasn't been derived yet. Even formal attempts at describing emergence still appeal to loose notions \autocite{Fletcher2021}. This issue of subjective delineations for emergent phenomena is a concern that has been identified before, but nonetheless is pervasive in the philosophical discourse on the topic \autocite{Taylor2015}.

\section*{Overview}

Our aim in this paper is to clarify some of these issues by clearly delineating different conceptions of emergence in ways that do not rely on personal judgments, and are explicit about the specific relationships between levels.
There have of course been numerous attempts at classifying different versions of emergence \autocite{sep-properties-emergent,wilson2021metaphysical}.
Our strategy has three particular features.
\begin{enumerate}
\item We take a physics-inspired approach by focusing not on properties or causal dependence, but rather on states and their dynamics.
We take a theory or model to consist of a specified space of states and some (possibly probabilistic) evolution rule, and investigating how lower-level theories can relate to higher-level ones in these terms.
    \item We pay explicit attention to the role of mereology, or the decomposition of systems into smaller constituent parts. In discussions of emergence it is typically assumed that macro systems can be thought of as being constituted by smaller pieces. Formally, this amounts to considering the state of a system as being expressible as a Cartesian product of the states of the individual constituents (which might be particles, or lattice sites, or people, or whatever). 
    We attempt to be clear about the role such a decomposition plays in emergence.
    \item We replace, insofar as is possible, notions of surprise and novelty with objective standards of the senses in which emergent theories might be described as new, unexpected, or difficult to foresee.
\end{enumerate}
We hope that this approach illuminates what would be required, at the level of states and dynamics, for different conceptions of emergence to pertain.

With these ideas in mind, we propose the following way of classifying varieties of emergence:
\begin{itemize}
    \item {Type-0 (featureless) emergence:} A many-to-one map from a micro theory without subsystems to a macro theory, which commutes with time evolution.
    \item {Type-1 (local) emergence:} A many-to-one map that commutes with time evolution where both theories describe collections of subsystems, and macro subsystems are made of specific micro subsystems.
    \begin{itemize}
    \item {Type-1a (direct) emergence:} Local emergence where the map is algorithmically simple.
    \item {Type-1b (incompressible) emergence:} Local emergence where the map is algorithmically complex.
    \end{itemize}
    \item {Type-2 (nonlocal) emergence:} Similar to Type-1, but allowing for macro entities or dynamics that are defined nonlocally from the micro perspective.
    \item {Type-3 (augmented) emergence:} A binary relation introducing new ontological features in the macro theory that are simply absent in the micro theory.
\end{itemize}
We do not put forward any one of our categories as capturing the uniquely correct formulation of emergence; we rather wish to help organize various conceptions that might be relevant to different contexts.

Our categories Type 0, 1, and 2 all correspond to different versions of what is traditionally labeled as weak emergence, as the higher-level dynamics are completely determined by those at the lower level.
The different types are distinguished by how direct and straightforward it is to deduce those higher-level behaviors.
Type-3 is close to strong emergence, in that it demands the introduction of new ontological entities at the higher level, which straightforwardly implies that lower-level dynamics is insufficient to predict the higher-level behavior.
The usefulness of our conception is that we can be explicit about what kind of theoretical structure is necessary to account for this situation.

How our categories relate to the epistemic/ontological distinction depends on one's thoughts about ontology.
Clearly, augmented (Type-3) emergence is an example of ontological emergence.
But as Dennett has argued, it can be completely appropriate to categorize the higher-level entities described by Type-1 and Type-2 emergence as ``real,'' in the sense of his ``real patterns" \autocite{Dennett1991-DENRP}.
The existence of an emergent description implies that it is possible to coarse-grain states of the micro theory and throw away an enormous amount of information about the microstate, and nevertheless retain somewhat accurate predictive power at the macro level.
The ability to make such predictions on the basis of so little data is a highly non-trivial fact about the micro theory, and the resulting coarse-grained structures seem well deserving of being labeled ``real."
(One crucial question we do not address is when such patterns exist, or how to quantify them \autocite{CRUTCHFIELD199411,shalizi2003macrostate,krakauer2020information,PhysRevE.108.014304,rupe2024principles}.)

\section*{Setup}

We imagine that there is a (presumably unknown) way of correctly and exactly describing the real world, and label that theory as $\Omega$.
It consists of a space of states $W$  with elements  $w \in W$.
If classical mechanics had turned out to be correct, the space of states would have the structure of phase space, consisting of positions and momenta for each of the components of the world.
Since classical mechanics is not correct, and we don't know the complete correct theory of the world, we will assume that there is some space of states, with specifying it further.

One important assumption that we do make is that there is evolution through time, and the relevant evolution law is Markovian, i.e. future evolution only depends on the current state (not, for example, on prior history that is not encoded in the current state).
This is a feature of all currently popular approaches to fundamental physics, but might seem to be less compelling at higher emergent levels.
Maybe some person's behavior is hard to understand without knowing some of their personal history.
But that can typically be accounted for by including memories and influences of that history as part of their current state.
We will therefore proceed with the Markovian assumption.
This means that states evolve with time under some evolution law:
\be
  w(t+\Delta t) = E_\Omega[w(t)]. 
\ee
In general the evolution law need not be deterministic, and we should speak about probability distributions rather than specific states. 
But this distinction won’t be relevant for our discussion, so to keep our notation relatively clean we will write as if everything is deterministic.

Now we imagine two theories that seek to  capture some aspects of  the world:
\begin{itemize}
    \item Theory $\alpha$ (``micro”), space of states $A$, individual states $a$, and evolution rule $a(t+\Delta t) = E_\alpha[a(t)]$.
\item Theory $\beta$ (``macro”), space of states $B$, individual states $b$, and evolution rule $b(t+\Delta t) = E_\beta[b(t)]$. 
\end{itemize}
For each theory, we again assume that the evolution is Markovian: the state at $t+\Delta t$ is determined by the state at $t$, without any additional knowledge of the prior history of the system. 

\begin{figure}[h]
\includegraphics[scale=0.7]{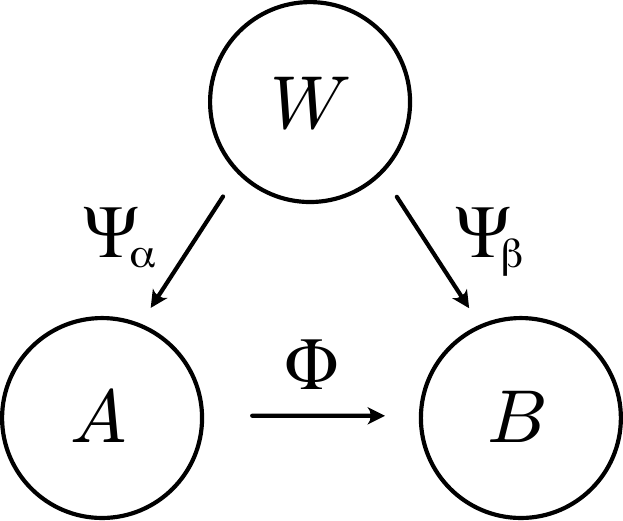}
\centering
\caption{Relations between the spaces of states of the world ($W$), the microscopic theory ($A$), and the macroscopic theory ($B$).}
\label{circles}
\end{figure}

The relationship between  states of  the world and  those of  these two theories can be illustrated in Figure~\ref{circles}.
In a slight abuse of notation, the arrows represent binary relations between subsets of two spaces of states.  
Arrows are natural to use because in many cases these relations take the form of  coarse-graining maps \autocite{list2019levels}, such that one or many states in the lower-level picture get mapped onto a single state of the higher-level picture (as when many states in a collection of atoms are described by the same distribution of density and other variables in a continuum description).  
But more carefully they should be thought of as binary relations rather than maps. 
Part of the definition of a map is that elements of the domain are associated with a unique element in the range, but the more general notion of a ``relation” simply associates elements of one set with those of another, even if non-uniquely. 
As we will see in the case of Type-3 emergence below, there could be examples where the ontology of $\beta$ includes features that are absent in  $\alpha$, so that a single microstate might correspond to multiple possible macrostates. 
The  relations might  also have some restricted domain, the ``domain of applicability” of the target theory (e.g. fluid mechanics only emerges from kinetic theory when there are sufficiently many particles).

We assume the relations are independent of time. 
Otherwise emergence would be trivial, as we could always adjust the  relations so as to  be compatible with time evolution. 
Note also that we will primarily be concerned about the ``emergence  relation'' $\Phi:A\rightarrow B$, but we indicate the separate  relations $\Psi_\alpha$ and $\Psi_\beta$ to emphasize that there might be states in  $W$  that are well-described by  $\beta$ but not by $\alpha$ (i.e., the domain of applicability of $\beta$  in  $W$  might not be a subset of the domain of applicability of  $\alpha$).

\section*{Type-0 (featureless) emergence}

\begin{figure}[h]
\includegraphics[scale=0.7]{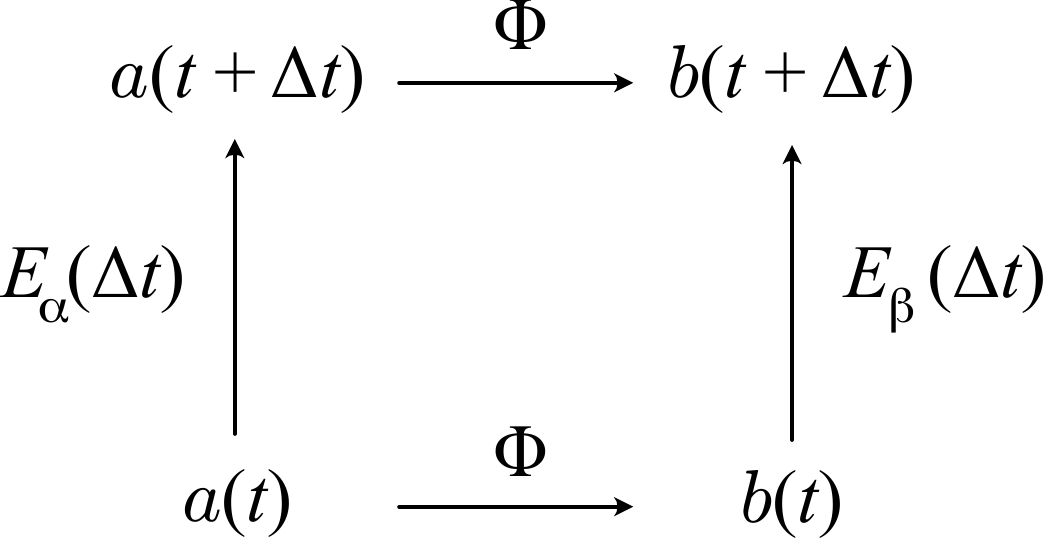}
\centering
\caption{The commuting diagram at the heart of emergence, relating time evolution and the emergence map.}
\label{commuting}
\end{figure}

In the first case we consider, the space of states  $A$ of the micro theory  $\alpha$  is not assumed to possess a preferred decomposition into subsystems.
The emergence relation $\Phi$ is a coarse-graining map, sending sets of microstates in $A$ to single macrostates in $B$. 
The criterion that this relation describes emergence is simply compatibility with time evolution, as expressed by the commutativity of the diagram in Figure~\ref{commuting}. 
Starting from some microstate $a\in A$ at time $t$, we can either act the emergence map to obtain a state $b\in B$ at $t$ and then evolve forward in time by $\Delta t$ according to the macro time-evolution operator $E_\beta$, or first time-evolve in the micro theory using $E_\alpha$ and then apply the emergence relation $\Phi$. 
Commutativity, the statement that we end up with the same final macrostate $b(t+\Delta t)$ either way, is the statement that the theory $\beta$ emerges from  $\alpha$.
Then we have:
\begin{itemize}
    \item \textbf{Type-0 (featureless) emergence:} 
    A map $\Phi$ (within a specified domain of applicability) from states $a\in A$ in a micro theory $\alpha$ to states $b\in B$ in a macro theory $\beta$, such that $\Phi$ commutes with the corresponding time-evolution operators $E_\alpha$ and $E_\beta$.
\end{itemize}

We do not insist that every state $a\in A$ in the micro theory have a corresponding macrostate $b\in B$; there may be a restricted domain of applicability $A_E \subseteq A$ where the emergent macro theory is accurate. 
For that matter, that accuracy may only be approximate rather than exact. 
It is typical of coarse-grained dynamics that they are only accurate up to some quantifiable level of precision, or with some quantifiable probability.

Type-0 emergence differs somewhat from conventional discussions of emergence, which take for granted a decomposition of the micro theory into subsystems, which then become ``parts'' of the macroscopic ``whole.''
But it shares the crucial feature of exhibiting efficacious dynamical laws even when microscopic information is discarded (coarse-grained over).
This is the essence of a real pattern.
It is not obvious that such effective macro dynamics should necessarily or typically exist; throwing away information about microstates generally makes accurate prediction impossible, so that this kind of emergence represents a special situation where very specific kinds of information can be ignored while maintaining macroscopic predictability. 

The idea that macro systems are made of parts described by microscopic dynamics is so prevalent that one might wonder whether there are any interesting examples of Type-0 emergence, but a moment's reflection reveals a paradigmatic case: quantum mechanics.
Quantum states are represented by vectors $|\Psi\rangle$ in Hilbert space $\mathcal{H}$, a complete normed complex vector space. 
For a collection of $n$ featureless non-relativistic particles, this state vector can be represented by an normalized complex-valued function of the particle positions, $\Psi(x_1, x_2, \ldots x_n)$.
The phenomenon of entanglement arises because this wave function will generally not be separable into a product of functions for each particle: $\Psi(x_1, x_2, \ldots x_n)\neq \Psi_1(x_1)\Psi_2(x_2)\cdots \Psi_n(x_n)$.
Therefore, the quantum system does not naturally decompose into distinct parts with independent dynamics; in general the entire wave function matters.\footnote{We may nevertheless decompose Hilbert space into subsystems, but entanglement implies that this will be given by a tensor product structure rather than the simpler Cartesian product. It then becomes an interesting question to ask how best to perform such a decomposition in different physical situations \autocite{Carroll:2020gme}.}

In certain special circumstances, however, Type-0 emergence occurs: namely, the classical limit.
If the wave function happens to be separable or nearly so, and the individual wave functions are relatively localized (and remain so, as will happen for high-mass particles) compared to distances over which the potential-energy function $V(x_1, x_2, \ldots x_n)$ varies, then Ehrenfest's Theorem implies that the expectation values of the positions and momenta, $\langle \hat{x}_i\rangle=\bra{\Psi}\hat{x}_i|\Psi\rangle$ and $\langle \hat{p}_i\rangle=\bra{\Psi}\hat{p}_i|\Psi\rangle$, approximately obey the classical equations of motion:
\be
 m_i \frac{d\langle\hat{x}_i\rangle}{dt} \approx \langle\hat{p}_i\rangle\, , \quad
 \frac{d\langle\hat{p}_i\rangle}{dt} \approx - \frac{\partial V}{\partial x_i}.
\ee
This is clearly a coarse-graining map: a wave function $\Psi(x)$, specified by a continuum of complex numbers, maps to a discrete set of positions and momenta.
Many distinct wave functions will map to the same classical position and momentum.
But the wave function doesn't have ``parts'' in the usual sense, so this is an example of Type-0 emergence.
Note the crucial role being played here by the restricted domain of applicability of the emergence map: classical mechanics only emerges from quantum mechanics under the right circumstances, such that entanglement and interference play no important role in the dynamics.

Other examples include other kinds of limiting behavior, such as the Newtonian limit of relativistic mechanics, when all relative velocities are much less than the speed of light.
There, within the domain of applicability, the relevant map is one-to-one rather than many-to-one (relativistic positions and momenta map to nonrelativistic equivalents), so this would be a somewhat trivial example of emergence.
There is also the Newtonian-gravity limit of general relativity, which requires both non-relativistic velocities and a weak gravitational field.
One could think of this as a many-to-one map, where the many macroscopically equivalent states of the micro theory (general relativity) are distinguished by negligible but nonzero amounts of gravitational radiation.

Type-0 emergence, based on nothing more than a commutative diagram featuring time evolution and the emergence map, is the basis for the more elaborate Type-1 and Type-2 emergence to be discussed below.
The defining feature of emergence is the existence of an effective macroscopic theory, where ``effective'' means that the dynamics of the macrostates can be described (at least to a good approximation) in terms of those macrostates alone, without reference to the lower-level micro-theory. 

This kind of commuting relation between two theories, while representative of distinctions in scientific disciplines, is also captured by other notions that relate system descriptions. 
Bickle's construction of functional, explanatory, and even mechanistic reduction can all be interpreted as various types of mappings through different procedures of course graining, all of which fit the Type-0 constraints \autocite{Bickle2024}.
Certain notions of computation also depend on commutativity between analytical descriptions and real phenomena \autocite{Horsman2014}. Thus, it is worth noting that while relations between scientific theories present the most clear example, type-0 and following emergence taxonomy will be applicable as broad meta-classifications for a myriad of theory-theory relations.  

In terms of the traditional strong/weak dichotomy, Type-0 is a subset of weak emergence.
Using Bedau's conception \autocite{Bedau1997-BEDWE}, if we knew the micro theory, the microstates, and the emergence map $\Phi: A\rightarrow B$, we could put the micro theory on a computer and infer the macro behavior.
Whether that behavior would be surprising or not is somewhat a matter of judgment, and we won't draw that distinction quite yet.

\section*{Subsystem Decomposition}

For Type-1 and higher versions of emergence, we consider cases where the micro and macro theories describe collections of interacting subsystems, and ask how those subsystems connect to each other under the emergence relation.

First we need to think about what it means to have a subsystem structure.
In many traditional examples, what counts as a subsystem is taken as given: one starts with a set of constituents, and collects them together to form a larger system.
But since notions of subsystems and locality will be crucial in what follows, it will pay to be specific about what these notions require and imply.

We will model a decomposition into subsystems as a Cartesian product.
That is, the space of states of the larger system is simply by an ordered sequence (or tuple) of states of all the subsystems.
Formally, for our lower-level theory $\alpha$ we have an injective map from the space of states $A$ into a Cartesian product of $I$ spaces $A_i$ of subsystem states $a_i$,
\be
  D_\alpha : A\rightarrow \prod_{i=1}^I A_i,
  \label{decomposition}
\ee
so that individual states are written as $I$-tuples,
\be
  D_\alpha : a \rightarrow (a_1, a_2, a_3, \ldots a_I).
\ee
In this notation, $a_i$ is the specific state of the $i$th subsystem, and so on.
A similar structure will hold for the higher-level theory $\beta$, which is decomposed into $N\leq I$ subsystems:
\be
D_\beta : B = \prod_{n=1}^N B_n, \qquad b \rightarrow (b_1, b_2, b_3, \ldots b_N).
\ee 
Cartesian products are not the only way to decompose a large system into subsystems; the notable example is quantum mechanics, where the overall system is described by the tensor product of the subsystems.
But other than the Type-0 case of quantum $\rightarrow$ classical emergence discussed above, systems of interest are generally decomposed into Cartesian products. 

However, there will generally be a very large number of conceivable decompositions of the form (\ref{decomposition}), so we need some way of picking the right one.
Our informal notion is that each subsystem should have a kind of identity of its own, and that the dynamics of the system as a whole can be thought of as described by the individual dynamics of the subsystems plus interactions between them. 
To make this somewhat more explicit, imagine writing the evolution law for each individual subsystem in the micro-theory as:
\be
  a_i(t+\Delta t) = E_{\alpha, i}^{(\mathrm{self})}[a_i(t)] 
  + E_{\alpha,i}^{(\mathrm{int})}[a_i(t), \{a_{\bar{j}(t)}\}]  .
  \label{subsystemdynamics}
\ee
Thus, the evolution of $a_i$ includes a self term for the subsystem under consideration, as well as an interaction term specifying the influence of other subsystems, with the notation $\{a_{\bar{j}}\}$ denoting the set of subsystems that interact directly with $a_i$.
The former depends only on the current subsystem state $a_i$, while the latter depends also on the state of other micro subsystems. 

The expression (\ref{subsystemdynamics}) is completely general; it merely reiterates the idea that the evolution of each subsystem depends on the state of that subsystem as well as that of all the other subsystems.
To pick out a useful decomposition, we need to implement the idea that the subsystems have individual identities.
This amounts to two requirements: (1) we can imagine turning off the interactions with other subsystems, either by ``moving them far away'' or by setting some coupling constants to zero, and (2) when those interactions are turned off, the data in the subsystem state $a_i(t)$ is sufficient to determine the evolution of that subsystem.
When these requirements are met for all the subsystems, we say we have an appropriate subsystem decomposition.\footnote{In practice we might also require some notion of maximality to our decomposition, so that no subsystem would be better thought of as the union of two smaller subsystems. But the appropriate notion of maximality will generally depend on other considerations (otherwise everything would be divided into elementary particles and fields), so we won't elaborate on that idea here.}

These requirements serve to exclude ways that in principle we could decompose a space of states into a Cartesian product, but where the factors fail to represent anything we would recognize as a subsystem.
For example, in the classical mechanics of a $n$ particles moving in $d$ spatial dimensions, where the state consists of the positions and momenta of all the particles, formally we could think of the set of all position variables as one subsystem and the set of all momenta as another.
But nobody is tempted to do that.
Mathematically we can perform such a factorization of phase space, but no particular insight is gained; moreover, knowing the position of a particle but not its momentum (or vice-versa) doesn't allow us to say anything about its future evolution.
Under the right conditions, however, it does make sense to treat position/momentum pairs $\{\vec{x}_i, \vec{p}_i\}$ as describing subsystems: to wit, individual particles.

A decomposition into subsystems allows us to think of the space of states as a network, with subsystems as nodes and interactions between subsystems represented by weighted edges.
In practice, interactions between subsystems will have different strengths, and some barely interact at all, so that there is a notion of which other subsystems are most relevant to the dynamics.
This hierarchy of dynamical relevance can be thought of as a notion of locality, with subsystems that interact noticeably thought of as ``near," and less relevant ones as ``far," although in higher-level theories those descriptions might not correspond to ordinary physical distance.

In lower-level models based on fundamental physics, spatial locality implies a natural notion of subsystems.
Consider for example the Ising model, consisting of spins defined on a lattice with direct interactions only with neighboring spins.
Then any particular spin is not directly affected by what is happening on distant lattice sites, only through the intermediaries of neighboring spins.
Classical field theories work similarly, where the dynamics of a field at one point in spacetime is directly affected only by the values of various fields (including itself), and their spacetime derivatives, at that same point and no others.
The derivatives are the continuum version of ``neighboring lattice sites,'' since they encapsulate infinitesimal changes from point to point.

But our notion of subsystems is more general than locality in space.
Consider a set of particles moving through space and interacting via Newtonian gravitation (an inverse-square law).\footnote{It is possible to define Newtonian gravity using a potential field obeying a differential equation, as shown by Pierre-Simon Laplace. But the field is not an independent dynamical degree of freedom, as it is completely determined by the distribution of matter (unlike in general relativity). So there is a sense in which the theory is nonlocal.}
Unlike with a lattice model, now the question of whether subsystems (the particles) are interacting strongly or not depends on the state (in particular, the distance between the particles).
There is nevertheless a sensible notion of subsystem individuality for particles, as we can always consider a decoupling regime for one particle where all other particles are far away and the trajectory of the particle under consideration depends only on its own state (and perhaps on a background gravitational field).
That stands in contrast with a misguided attempt to think of position and momentum as subsystems; without knowing the momentum of a particle, we can't predict its subsequent position at all. 

As a final example, take a network of people interacting over the internet.
Individuals will directly interact with some other individuals but not others, but the strength of those interactions is not directly tied to physical distance.
A useful decomposition into subsystems, in other words, plays nicely with the dynamics, giving some notion of autonomy to each subsystem under the right conditions.

\section*{Type-1 (local) emergence}

With that in mind, we define:
\begin{itemize}
    \item \textbf{Type-1 (local) emergence:} Type-0 emergence  (i.e. there exists a map $\Phi:A \rightarrow B$ that commutes with time evolution), with the additional condition that subsystems in the macro theory consist of localized collections of subsystems in the micro theory. 
\end{itemize}
``Consist of'' here means that the state $b_n$ of each specific subsystem $B_n$ in the emergent theory depends solely on the states $\{a_i^{(j)}\}$ of a specified collection of subsystems of the micro theory $\{A_i^{(j)}\}$, such that each micro subsystem only contributes to the state of one macro subsystem.

Type-1 includes many standard examples of weak emergence, especially in physics.
Consider the Ising model mentioned above, or analogous lattice models with nearest-neighbor interactions. 
As suggested by Kadanoff in early discussions of the renormalization group, we can partition the original lattice into $n\times n$ regions, and associate with each region a ``block spin'' given by the average of the actual spins.
We can then discuss the dynamics of the theory in the long-distance limit by referring only to the block spins and their interactions, in a direct example of emergence \autocite{kadanoff1966scaling}.
More generally, effective quantum field theories with an ultraviolet energy cutoff can be thought of as locally-emergent theories valid below the cutoff \autocite{burgess2007introduction}.

Another example is provided by center-of-mass motion in classical Newtonian mechanics.
In order to predict the motion of the Earth around the Sun, we needn't know the state of every particle constituting the Earth; we need only know the center-of-mass coordinates and momenta of the Earth and other solar system bodies.
Indeed, the center-of-mass example was used by Dennett as a paradigmatic instance of a real pattern \autocite{Dennett1991-DENRP}.
From our perspective, this reflects the fact that we can make accurate predictions for the trajectories of celestial bodies even after coarse-graining their microstates to just the center-of-mass data, thereby ignoring the vast majority of information needed to characterize the exact microstate. 
It is crucial that we coarse-grain in the right way; throwing away arbitrary subsets of the micro data would leave us completely unable to make reliable predictions.

In the case of the Ising model or center-of-mass motion, the basic ontology of the macro theory is structurally the same as that of the micro theory.
There is nothing in our definition of Type-1 emergence that demands this.
Another common example of emergence is the relationship between a microscopic theory of many interacting particles and a macroscopic fluid description.
In that case the emergence map takes a small volume of space and calculates appropriate averages over the states of the particles within that volume to determine macroscopic quantities like pressure and temperature, which can be treated as a continuum in the limit of taking the averaging volume to be small (but large enough to contain many particles).
A fluid is ontologically different from a set of discrete particles, but this is nevertheless a straightforward example of Type-1 emergence.

One subtlety here is that \emph{which} micro subsystems contribute to the state of a specific macro subsystem may (or may not) depend on the overall state of the micro theory.
In the particles-and-fluids case, for example, individual particles (subsystems from the micro perspective) will move from one small averaging volume into others.
This is fine, as it doesn't conflict with the idea that the macro subsystem is made out of particular micro subsystems.
Similarly, a celebrated example of emergence is Conway's Game of Life, a two-dimensional cellular automaton featuring collective structures such as blinkers and gliders \autocite{izhikevich2015game,Bedau1997-BEDWE}.
Since gliders move, changing which set of cells they occupy, the micro systems (lattice sites) that make them up will also change, but that is consistent with our definitions.
The Game of Life is also an example of how the macro theory can exhibit complex behavior when the micro rules are relatively simple, although that phenomenon doesn't play any central role in how we have characterized emergence here.

Defined in this way, either Type-0 or Type-1 emergence falls under most traditional notions of weak emergence: we could take the state and evolution rules of the micro theory, put it on a computer, and get behavior that is compatible with the prediction of  the macro theory. 
We nevertheless think the distinction is worth drawing, since what we have called Type-0 is not always recognized as emergence at all, and explicitly highlighting the role of a local decomposition will be important when considering further varieties of emergence.

\section*{Novelty and compressibility}

Despite the straightforwardness of Type-1 emergence, there can be cases falling under this category in which the emergent theory and its properties might nevertheless be plausibly described as ``novel” or ``surprising” or ``unexpected” or even ``unpredictable'' \autocite{anderson1972more}.
But these words seem to be about human judgments rather than objective qualities.
What if someone fails to be surprised, or claims to have expected the emergent behavior all along?
Couldn't any behavior in a locally-emergent macro theory be predictable, in principle, by first following the dynamics in the micro theory and then applying the emergence map?

We can try to capture some of this feeling in a more objective way by considering properties of the actual emergence map: in particular, whether it is simple or complex. 
We are thinking here of algorithmic complexity in the sense of Kolmogorov, Chaitin, and others: the length of the shortest computer program that will produce a desired output \autocite{li2008introduction}.
When there is a compact formula that produces a desired output (such as a string of one billion zeros), the output is said to be simple or compressible; when the shortest program contains an explicit representation of the output and simply prints it (such as a specific billion-digit number with no special properties), the output is complex or incompressible.

In our case, the algorithm in question is the emergence map $\Phi:A \rightarrow B$.
That map, typically many-to-one, may be expressible in terms of a compact equation, or it might require explicit delineation of which microstates get mapped to which macrostates (or various degrees in between).
In the case of center-of-mass motion or particles-to-fluids emergence, for example, it is straightforward to give explicit formulae that construct macrostates from microstates; the mass density is simply the number of particles per volume times their individual masses, the pressure is related to the kinetic energies of the particles, and so on.
Whereas in the Game of Life, the definition of a glider cannot be made much more compact than an explicit representation of the appropriate states of the individual cells.  
This leads to the following sub-classifications of local emergence:
\begin{itemize}
    \item \textbf{Type-1a  (``direct”)  emergence:} Type-1 emergence based on a compressible map $\Phi$.
    \item \textbf{Type-1b (``incompressible”)  emergence:} Type-1 emergence based on an incompressible map $\Phi$.
\end{itemize}
Admittedly, this distinction is a spectrum rather than a clear-cut binary; some maps may fall in between extreme simplicity and utter incompressibility.
But the level of compressibility is at least an objective feature of the map, rather than an expression of our attitudes toward it.\footnote{In general, algorithmic compressibility cannot be computed, but it can be approximated \autocite{jansen2006introduction}.}

The direct/incompressible distinction is meant to capture some of what is meant by the novelty or unexpectedness of emergence.
When the emergence map is direct, we tend to think of the resulting emergent behavior as unsurprising; the existence of a simple map tracks our intuitive feeling for what we should expect from collective behavior.
The ontology of fluid mechanics may be different from that of kinetic theory, but the physical behavior of fluids makes sense to us once we learn that fluids are made of molecules.
Whereas the need for an explicit, incompressible emergence map tends to go hand-in-hand with emergent behavior being surprising or unexpected.
It's not that most collections of micro subsystems would behave in some particular macroscopic way, but that very particular ones do.
That kind of behavior is harder to intuitively perceive from the macro point of view.

\section*{Type-2 (nonlocal) emergence}

The next type of emergence we consider is a different refinement of Type-0, one where locality at the micro level is not manifest at the macro level. 
\begin{itemize}
    \item \textbf{Type-2 (nonlocal) emergence:} Type-0 emergence where micro and macro theories have decompositions into local subsystems, but where the macro theory does not respect the notion of locality inherited from the micro theory.
\end{itemize}
There are basically two ways this can happen.
First, the macro theory could involve degrees of freedom that are defined, in principle, in terms of the micro state, but they are not made of local collections of micro entities -- they are defined globally, in terms of many dispersed parts.
Then the macro entities can’t be thought of as simply bundles of nearby subsystems acting in a coherent fashion. 
Second, the macro entities could interact with each other in ways that would seem nonlocal from the micro point of view.
In either case we need to know the states of many (or even all) micro subsystems in order to successfully predict the macro behavior – there is emergence, but not independent emergence of distinct groups of subsystems.
We will not introduce separate subcategories for these two possibilities, as in practice they will generally appear together.
(The direct/incompressible distinction that we introduced for Type-1 emergence can also be considered in the context of Type-2, but we won't introduce explicit subcategories for this, either.)

This type of emergence generally doesn't happen when the macro theory is not too far removed from the particles and fields of fundamental physics.
In such cases, the strict locality of the micro theory tends to be directly carried over to the macro theory, placing limits on the existence of truly nonlocal entities even at the emergent level.
In a ferromagnet, for example, there can be nonlocal-seeming \emph{behavior}, when interactions between spins lead to an overall direction of magnetization that is shared throughout the material.
This is an example of spontaneous symmetry breaking, where individual solutions to the equations do not respect the full set of symmetries of the equations themselves.
But the dynamics remains local, even at the emergent level.

Nonlocal emergence can become relevant, however, in contexts like biology or social sciences.
The underlying limit on information propagation imposed by the speed of light is still present, but becomes largely irrelevant, essentially because of the timescales involved.
For all practical purposes, information can be shared across distances much faster than the dynamical timescales describing the motions of macroscopic entities, so there is no inconsistency in having emergent entities that are not effectively localized in space.
In psychology and cognitive science, for example, our higher-level theory of human beings might contain things like mental states, which cannot be associated with localized groups of neurons or cells, much less with atoms or particles.
(They can be associated with appropriate states of \emph{all} the neurons, or all the particles, just not with localized groups of them.) 
Examples would include Searle's biological naturalism \autocite{Searle2007-SEABN-2} and the Integrated Information Theory approach to consciousness \autocite{Tononi2016}.
Likewise if some sort of social or group dynamics are the higher-level theory, it is natural to include global variables as part of the most parsimonious emergent description.
A treaty agreement might play an important role in a causal theory of international relations, but as an emergent entity it can't be identified with a particular collection of micro entities.


By our definition of Type-2 emergence, there can be nonlocal variables in the macro theory, but they are ultimately determined by the state of the micro theory.
In that sense they are not truly new.
Type-2 does, nevertheless, open the possibility that nonlocal effects feed back onto the micro dynamics in ways that might not have been obvious from a purely micro perspective.
It can therefore seem as if the macro theory exhibits behaviors that the micro theory cannot account for, but can ultimately be understood by saying that the micro theory was simply incomplete, rather than incorrect.

Consider the following possible evolution law for a micro subsystem $a_i(t)$:
\be
  a_i(t+\Delta t) = E_{\alpha, i}^{(\mathrm{self})}[a_i(t)] 
  + E_{\alpha,i}^{(\mathrm{int})}[a_i(t), \{a_{\bar{j}(t)}\}] 
  + \sigma_i[a(t)] E_{\alpha,i}^{(\mathrm{nl})}[a_i(t),\{a_{\bar{j}'(t)}\}] .
  \label{nonlocaldynamics}
\ee
The first two terms are conventional, and are familiar from (\ref{subsystemdynamics}): the evolution of $a_i$ includes a self term for the subsystem under consideration, as well as an interaction term specifying the influence of other subsystems, with the notation $\{a_{\bar{j}}\}$ denoting the set of subsystems that interact directly with $a_i$.
The former depends only on the current subsystem state $a_i$, while the latter depends also on the state of other micro subsystems (with less influence from subsystems that are farther away).

The third term is the new contribution allowed in Type-2 emergence.
It includes a new nonlocal interaction factor $E_{\alpha,i}^{(\mathrm{nl})}$ that depends on both the subsystem itself and a (potentially different) set $\{a_{\bar{j}'}\}$ of other subsystems. 
This novel term is multiplied by a filter function $\sigma_i[a(t)]$ that depends on the global micro state.
The filter function goes to zero in situations that are exclusively in the micro domain, such as when a small number of particles are interacting in a physics experiment.
But it becomes nonzero in special circumstances that can depend on the global state of the system, perhaps in subtle ways.

In other words, in what we are calling Type-2 emergence, it could indeed be true that electrons behave differently inside a human brain than would be predicted by quantum field theory as we know it. 
The filter function $\sigma_i$ might turn on when atoms are arranged brain-wise, and remain zero otherwise.
But in laboratory experiments involving relatively small numbers of constituents, including high-energy collisions at particle accelerators, their behavior could remain correctly described by conventional QFT to very high accuracy, essentially using just the first two terms in (\ref{nonlocaldynamics}).
The difference is (or might conceivably be, at any rate) that the dynamics is affected in ways that are undetectable in a variety of conventional experimental conditions.
Particle-physics experiments are, quite sensibly, usually performed on small numbers of particles at a time.

To be clear, everything we know about quantum field theory assures us that this doesn't happen: the knowledge we gain about particle interactions from such experiments should allow us to make accurate predictions even when many particles come together in brain-like configurations \autocite{carroll2021consciousness,carroll2022quantum}.
But perhaps what we think we know about QFT is wrong in these situations.
There is no principled reason to suspect this from a purely particle-physics perspective, but it is a logical possibility.

But there are other cases where Type-2 emergence is perfectly compatible with physics as we know it.
DeDeo \autocite{DeDeo_2018} highlights the example of the ``jerk'' -- the derivative of the acceleration with respect to time (i.e., the third derivative of position).
This is not part of the fundamental ontology of classical mechanics, which is defined in terms of positions and velocities (or momenta) of the relevant degrees of freedom.
Acceleration, the first derivative of velocity, can be calculated in terms of the instantaneous state of the system ($\vec a = \vec{F}/m$), but jerk cannot; one would have to follow the system for some bit of time.
Nevertheless, we experience the jerk in our everyday lives, e.g. in the motion of a car or an elevator.
This is a case of Type-2 emergence, with the lower-level theory given by classical mechanics and the higher-level theory given by human-level experience.
One way of thinking about it is that the jerk is not calculable in terms of local and instantaneous quantities of the classical state, but could be calculated if one were to coarse-grain in time as well as in space.
However, that is not strictly necessary; information contained in the \emph{global} instantaneous state suffices, according to the rules of classical physics, to determine all of the future evolution of the system.
In our setup, the jerk experienced by a macroscopic person supervenes on some not-entirely-local set of micro data (e.g. the state of a car and the driver and the road conditions, as well as the local state of the passenger).
In relativistic mechanics the relevant amount of non-local information is restricted to the interior of the past light cone of the subsystem in question, but even in non-relativistic mechanics there will be an effective set of not-too-far-away influences that are relevant to determining the jerk.

Discussions of emergence frequently touch on the possibility of downward causation: higher-level entities exerting causal influence on lower-level entities. 
Strictly speaking, such a phenomenon is incompatible with either Type-1 or Type-2 emergence as we have defined it; in both cases the dynamics of the micro subsystems are fully determined in their own right. 
But the possibility of novel macroscopic interactions such as in (\ref{nonlocaldynamics}) in Type-2 emergence can lead to a kind of counterfeit downward causation.
To a macro observer, it might appear as if higher-level features are directly influencing behaviors of the micro systems, even though in principle the micro dynamics are entirely self-contained.

\section*{Type-3 (augmented) emergence}

All of the varieties of emergence we have thus far considered would traditionally be classified as weak emergence under most definitions thereof.
In each case, the higher-level dynamics supervene on the lower-level degrees of freedom.
The connection between the levels may be straightforward or it may be subtle, but it is nevertheless ironclad.

Strong emergence is supposed to be different.
It posits macroscopic effects that cannot be thought of, even in principle, as simply arising from the collective behavior of smaller parts obeying the microscopic dynamics.
The whole is truly different than the sum of its parts.

Strong emergence is most often taken seriously in approaches of consciousness, for example Chalmers's solution to the hard problem or theories of emergent panpsychism \autocite{Chalmers2007, sep-panpsychism}.  
In our terminology, these theories attempt to describe a macro model involving consciousness that stands as strongly emergent with respect to a micro model of the world (for the aforementioned theories, this tends to be the traditional physical/biological model of reality and conscious entities). 
In these theories of consciousness, some form of Type-3 relation between the model that involves consciousness and the physical or biological micro model exists. 
A new kind of object or entity, independent of and bearing no relation to the decomposition of the micro theory, is put forth as crucial to consciousness. 

But the idea of strong emergence is challenging to make sense of.
If a higher-level system is a collection of lower-level parts, and we have a dynamical micro theory that accurately describes how those parts behave, then that micro theory will make specific predictions about how the collection will behave.
Such predictions are either going to be correct, or incorrect.
If they are incorrect, it makes sense to say that the micro theory was just wrong from the start; if they are correct, any macro behaviors should in principle be reducible to the predictions of the micro theory.
There seems to be little room for both the micro theory to be correct, and for any new macro behavior we would classify as strongly emergent.
In other words, causal closure of the micro theory is incompatible with the kinds of new macro entities required by strong emergence \autocite{Kim2006-KIMECI}.

Here we propose a way, within the framework we've been considering, to have a version of strong emergence that avoids this apparent inconsistency.
The basic idea is to imagine that there are entities in the ontology of the macro theory that do not supervene on micro subsystems, but whose dynamical relevance disappears in the limit where we consider small numbers of microscopic systems.
Then one theory is not derivable from the other, but both represent what a reasonable scientist might develop to describe states and dynamics in the appropriate regime.
We refer to this possibility as ``augmented'' emergence, due to the appearance of these additional macro entities.
In this way, the micro theory could be used to make predictions for macro questions, and those predictions would (at least some of the time) turn out to be incorrect, even though the micro theory is perfectly accurate when deployed purely in the micro regime.

To be more specific, we imagine having a micro theory $\alpha$ with a decomposition of states into local subsystems, $A= \prod_i A_i$. 
This theory is successful as long as we consider situations in which only a few subsystems at a time are evolving and interacting. 
And there is a macro theory $\beta$ that is successful as well, but in a domain that (from the point of view of the micro theory) involves a very large number of mutually-interacting subsystems (at least for certain states). 
Because these two domains are distinct, it is possible that there are parts of the ontology of $\beta$ that have no counterpart in theory $\alpha$. 
The space of states of the macro theory is then spanned by both a set of supervenient variables $\{b_n\}$ that are uniquely determined by the micro variables, and a new set of independent variables $\{\bar{b}_q\}$ that are not.
In that case, the relation $\Phi$ associating states in  $A$  with those in $B$ need not be a well-defined map, in the sense that single elements of  $A$  might correspond to multiple elements of $B$ (ones with different values of the $\{\bar{b}_q\}$). 
Rather, it can be thought of as a binary relation, connecting microstates to multiple macrostates with different values of the nonlocal variables.
We can then define:
\begin{itemize}
    \item \textbf{Type-3 (augmented) emergence:} A macro theory featuring both variables that are constructed from states of the micro theory, and new variables that are independent of micro states, but which become dynamically relevant only in certain macroscopic situations.
\end{itemize}
This corresponds to what is conventionally thought of as strong emergence. The macro theory does not reduce to the micro theory, but the micro theory is not simply wrong; it just doesn't apply in the regime where the macro theory would most naturally be used.

We can think of the dynamics in theories of augmented emergence similarly to the nonlocal Type-2 effects sketched in (\ref{nonlocaldynamics}).
In Type-3, we can express the evolution law for micro subsystems $a_i$ (such as some particular electron) as follows:
\be
  a_i(t+\Delta t) = E_{\alpha, i}^{(\mathrm{self})}[a_i(t)] 
  + E_{\alpha,i}^{(\mathrm{int})}[a_i(t), \{a_{\bar j}(t)\}] 
  + \sigma_i[a(t)] E_{\alpha,i}^{(\mathrm{aug})}[a_i(t), \{a_{\bar{j}'}(t)\}, \{\bar{b}_q(t)\}] .
  \label{augmenteddynamics}
\ee
As in the Type-2 case, the first two expressions are conventional self and interaction terms.
The third term is the new contribution in Type-3 emergence.
It includes an interaction factor $E_{\alpha,i}^{(\mathrm{aug})}$ that couples the micro subsystem $a_i$ to the novel macroscopic variables $\{\bar{b}_q\}$, as well as to a possible set of other micro subsystems.
The filter function $\sigma_i[a(t)]$ once again goes to zero in situations that are exclusively in the micro domain and becomes nonzero in appropriate circumstances, depending on the global state.
In that way, the new augmented variables $\{\bar{b}_q\}$ only become dynamically relevant in those particular circumstances, and could remain invisible to experiments with just a few moving parts.

Once again, from the point of view of physics as it is currently understood, there is no reason at all to suspect something like (\ref{augmenteddynamics}) to be true, and many reasons to doubt it, at least when the microscopic variables are fundamental particles and fields.
Any full-blown theory along these lines would have to establish that the consequent dynamics are well-defined, that there are no runaway instabilities (ghosts), that gauge invariance and conservation laws are respected (or at least violated only at experimentally permitted levels), and more.

But problems such as consciousness, and perhaps the origin and evolution of life, are difficult, and it is useful to have an explicit framework with which to think about deviations from what we think we know about physics.
And something like augmented emergence might be more natural in circumstances where the subsystems of the micro theory are themselves complex, from biological cells to human agents.

\section*{Conclusion}

We offer this contribution to an already-considerable literature on emergence in order to help clarify the relationship between philosophical discussions of emergence and a physics-oriented view of states and their dynamics. 
We take the basic feature of emergence to be the existence of a particular kind of real pattern -- a coarse-grained macroscopic theory that is able to make accurate predictions on the basis of much less data than is contained in the microscopic description.
By carefully considering the decomposition of systems into subsystems with certain types of interactions, and examining different relationships between such decompositions at the micro and macro level, we are able to classify varieties of emergence explicitly and without leaning on subjective characterizations.

\section*{Acknowledgments} 

We are very grateful to a number of generous readers provided invaluable suggestions on the manuscript, including David Chalmers, Simon DeDeo, David Kinney, David Krakauer, and Elanor Taylor.
We thank Tyler Milhouse for his encouragement and patience.
And of course we offer thanks and appreciation to Dan Dennett, for inspiration and influence both evident and subtle.

\bibliographystyle{utphys}
\bibliography{references}
\end{document}